\documentclass[prb,12pt,a4paper,amsmath,amssymb]{revtex4}

\usepackage{color,epsfig,rotating,graphicx,subfigure}
\usepackage{amsmath,amssymb,amscd}

\usepackage[latin1]{inputenc}
\usepackage[english]{babel}

\usepackage{dsfont}
\usepackage{textcomp}
\renewcommand{\Im}{{\rm Im}}

\newcommand{\re}{{\rm e}}
\newcommand{\rd}{{\rm d}}

\newcommand{\rs}{{\rm s}}
\newcommand{\rp}{{\rm p}}

\newcommand{\kb}{k_{\rm B}}

\begin{document}
\title{Super-Planckian Near-Field Thermal Emission with Phonon-Polaritonic Hyperbolic Metamaterials}

\author{S.-A. Biehs and M. Tschikin}

\affiliation{Institut f\"{u}r Physik, Carl von Ossietzky Universit\"{a}t,
D-26111 Oldenburg, Germany.}

\author{R. Messina and P. Ben-Abdallah}

\affiliation{Laboratoire Charles Fabry, UMR 8501, Institut d'Optique, CNRS, Universit\'{e} Paris-Sud 11, 2, Avenue Augustin Fresnel, 91127 Palaiseau Cedex, France.}

\date{\today}


\begin{abstract}
We study super-Planckian near-field heat exchanges for multilayer hyperbolic metamaterials using 
exact S-matrix calculations. We investigate heat exchanges between two multilayer
hyperbolic metamaterial structures. We show that the super-Planckian emission of such 
metamaterials can either come from the presence of surface phonon-polaritons modes or 
from a continuum of hyperbolic modes depending on the choice of composite materials 
as well as the structural configuration.
\end{abstract}

\maketitle

%%%%%%%%%%%%%%%%%%%%%%%%%%%%%%%%%%%%%%%%%%%%%%%%%%%%%%%%%%%%%%%%%%%%%%%%%%%%%%%%
%
% Introduction
%
%%%%%%%%%%%%%%%%%%%%%%%%%%%%%%%%%%%%%%%%%%%%%%%%%%%%%%%%%%%%%%%%%%%%%%%%%%%%%%%%
In the last few years several fascinating experiments have demonstrated that for 
small separation distances compared with the thermal wavelength the thermal radiation 
exchanged between two hot bodies out of thermal equilibrium increases dramatically compared 
with what we observe at large distances and can even exceed the well-known Stefan-Boltzmann 
law by orders of magnitude~\cite{HuEtAl2008,NanolettArvind,ShenEtAl2008,NatureEmmanuel,Ottens2011,Kralik2012}.
Accordingly, thermal emission is in that case also called super-Planckian emission
emphasizing the possibility to go beyond the classical black-body theory. There are several promising 
applications of super-Planckian emitters ranging from thermal imaging~\cite{Yannick,KittelEtAl2008,HuthEtAl2011} and thermal 
rectification/management~\cite{Fan2010,FrancoeurRect,BiehsEtAl2011APL} to near-field 
thermophotovoltaics~\cite{MatteoEtAl2001,NarayanaswamyChen2003,LarocheEtAl06,ParkEtAl2007,ZhangReview}. 
This has triggered many studies on the possibilities of tailoring and controlling the super-Planckian 
radiation spectrum by means of designing the material properties~\cite{Zhang2005,Joulain2010,BiehsEtAl2011,Guerot2012,LussangeEtAl2012,Cui2013}, using phase-change materials~\cite{Zwol2010b} or 2D systems such as 
graphene~\cite{Svetovoy2011,IlicEtAl2012}, for instance. 
 
Recently, it was shown that hyperbolic metamaterials can lead to broad-band photonic thermal
conductance inside the material itself~\cite{NarimanovSmolyaninov2011} and between two hyperbolic materials
only separated by a vacuum gap~\cite{BiehsEtAl2012}. Further Nefedov {\itshape et al.} considered
nanorod-like structures made of nanotubes which are interlocked and highlighted a giant radiative
heat flux which could be utilized for near-field thermophotovoltaics energy conversion~\cite{Nefedov2011}. Finally,
Guo {\itshape et al.} have studied the energy density produced by the thermally fluctuating
fields close to a hyperbolic structure and found a broadband near-field contribution
from which they have concluded that super-Planckian emission will be broad-band for
hyperbolic materials~\cite{GuoEtAl2012}. This is in accordance with the findings in Ref.~\cite{BiehsEtAl2012}
for the energy exchange between two hyperbolic nanowire structures. 

The aim of this letter is to show that the surface modes supported by the topmost layers of 
phonon-polaritonic metamaterials can give the dominant contribution to the super-Planckian emission. 
As was shown in Ref.~\cite{TschikinEtAl2013} materials which have a broad hyperbolic frequency 
band as predicted from effective medium theory can support surface modes inside these frequency 
bands as well which will compete with the hyperbolic modes~\cite{Rosenblatt2011,TschikinEtAl2013}. 
In particular, we will show that for the realization of a hyperbolic metamaterial as studied 
in Ref.~\cite{GuoEtAl2012} the main contribution to super-Planckian radiation is not necessarily due to hyperbolic 
modes but can be due to surface modes depending on the choice of the topmost layer. We will show
that in order to allow for broad-band super-Planckian emission by hyperbolic modes, mainly, it is 
important to use a material for that topmost layer which does not support surface modes 
in the thermal freqeuency range. 

Before we start to study the super-Planckian thermal radiation let us first recall the concept
of indefinite or hyperbolic materials~\cite{Smith2000,SmithSchurig2003,HuChui2002}. Such materials 
are first of all a special class of uni-axial anisotropic materials. For uni-axial materials the 
permittivity $\epsilon_\parallel$ parallel to the optical axis is different from the permittivity 
$\epsilon_\perp$ perpendicular to the optical axis. For hyperbolic materials one can find frequency 
bands where $\epsilon_\parallel$ and $\epsilon_\perp$ have different signs, i.e.\ $\epsilon_\perp \epsilon_\parallel < 0$. 
Thus the dispersion relation of the photons in such a material~\cite{Yeh}
\begin{equation}
    \frac{\kappa^2}{\epsilon_\perp} + \frac{k_{z}^2}{\epsilon_\parallel} = \frac{\omega^2}{c^2}
\label{Eq:HyperbolicDisp}
\end{equation}
describes a hyperbolic function rather than an ellipse as for usual anisotropic materials; here
$\kappa$ ($k_z$) is the wavevector inside the hyperbolic medium perpendicular (parallel) to the optical axis which is assumed
to point in $z$-direction. Such metamaterials can for example be designed by multilayer structures, 
since in the long wavelength regime such structures can be described as homogeneous anisotropic 
media with the effective permittivities 
\begin{align}
  \epsilon_\perp &= \epsilon_1 f + \epsilon_2 (1 - f), \label{Eq:epsperp}\\
  \epsilon_\parallel &= \frac{1}{f/\epsilon_1 + (1-f)/\epsilon_2}, \label{Eq:epsparallel}
\end{align}
where $\epsilon_1$ and $\epsilon_2$ are the permittivities of the two layer-materials and $f$ is the
filling fraction of the topmost material $1$, i.e.\ $f = l_1/(l_1 + l_2)$. The effective permittivities allow for a calculation of the hyperbolic 
frequency bands of the multilayer structure where $\epsilon_\perp \epsilon_\parallel < 0$. There are
in general two different kinds of bands: a frequency band $\Delta_1$ where $\epsilon_\parallel < 0$
and $\epsilon_\perp > 0$ and a frequency band $\Delta_2$ where $\epsilon_\parallel > 0$
and $\epsilon_\perp < 0$. As will become clear in the following the such calculated frequency 
bands $\Delta_1$ and $\Delta_2$ can also support surface modes, which are not taken into account 
in the effective description~\cite{TschikinEtAl2013}.

\begin{figure}[Hhbt]
  \centering
  \epsfig{file=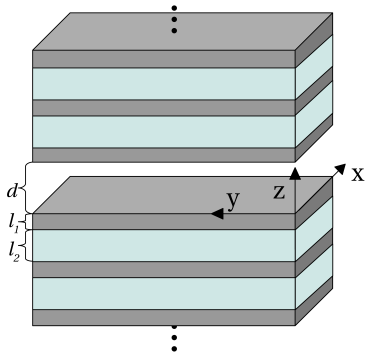, height = 0.3\textwidth}
  \caption{\label{Fig:Structure} Sketch of the geometry of two hyperbolic multilayer materials separated by a vacuum gap.}
\end{figure}

In order to study super-Planckian radiation we consider the geometry depicted in Fig.~\ref{Fig:Structure}.
The heat transfer coefficient $h(d)$ between the two metamaterials which are assumed to be at local 
thermal equilibrium can be determined by~\cite{SurfaceScienceReports}
\begin{equation}  
  h(d) = \int_0^\infty\!\!\frac{\rd \omega}{2 \pi} f(\omega,T) \sum_{j = \rs,\rp} \int\!\!\frac{\rd^2 \kappa}{(2 \pi)^2} \mathcal{T}_j(\omega,\kappa;d) = \int_0^\infty\!\!\frac{\rd \omega}{2 \pi} f(\omega, T) H(\omega,d)
\label{Eq:htc}
\end{equation}
where $f(\omega,T) = (\hbar \omega)^2/(\kb T^2)\re^{\hbar \omega/\kb T}/(\re^{\hbar \omega/\kb T} - 1)^2$.
$\mathcal{T}_\rs(\omega,\kappa;d)$ and $\mathcal{T}_\rp(\omega,\kappa;d)$ are the energy transmission coefficients
for the s- and p-polarized modes which can be easily determined for semi-infinite materials, 
anisotropic materials and multilayer structures~\cite{Biehs2007,FrancoeurAPL,LauEtAl2008,PBA2009,Francoeur2009,PBA2010,BiehsEtAl2011,Pryamikov2011,TschikinEtAl2012,MessinaEtAl2012,MaslovskiEtAl2013}. Here we use the standard S-matrix approach 
as in Refs.~\cite{Francoeur2009,TschikinEtAl2012} to calculate the amplitude reflection coefficients 
$r_j$ of our multilayer structures from which we can easily determine the energy transmission coefficients
\begin{equation}
   \mathcal{T}_{j}(\omega,\kappa; d) =
    \begin{cases}
     (1 - |r_j|^2)^2 / |D_j|^2, & \kappa < \omega/c\\
     4 [\Im(r_j)]^2{\rm e}^{-2 |k_{z0}| d}/|D_j|^2 ,  & \kappa > \omega/c
  \end{cases}
\label{Eq:TransmissionCoeff}
\end{equation}
including the contributions of the propagating modes with $\kappa < \omega/c$ and the evanescent 
modes with $\kappa > \omega/c$. Here $D_j = 1 - r_j r_j {\rm e}^{2 {\rm i} k_{z0} d}$
is a Fabry-P\'{e}rot-like denominator with $k_{z0}^2 = k_0^2 - \kappa^2$; $k_0 = \omega/c$. 

Now, let us consider a concrete example of a hyperbolic structure which is composed by layers of polar materials. 
Because these structures can support surface phonon-polaritons as well, they are also called 
phonon-polaritonic hyperbolic structures. We choose to consider the structure
in Ref.~\cite{GuoEtAl2012} which is made of layers of SiC and SiO$_2$. In general amorphous SiO$_2$ supports
surface modes in the infrared as well as SiC, but to get results which are comparable with the calculations
done in Ref.~\cite{GuoEtAl2012} we assume that $\epsilon_{\rm SiO_2} = 3.9$ adding a vanishingly small absorption.
The optical properties of SiC are taken from Ref.~\cite{Palik}. The layer thicknesses are (a) $l_1 = 50\,{\rm nm}$ 
for the SiC layers and $l_2 = 150\,{\rm nm}$ for the silica layers so that the filling fraction is $f = 0.25$
and (b) $l_1 = l_2 = 100\,{\rm nm}$ so that $f = 0.5$. For our exact S-matrix calculations
we use $N = 50$ where the last layer is a semi-infinite layer with the material properties of the topmost layer. 
The hyperbolic frequency bands calculated from Eqs.~(\ref{Eq:epsperp}) and 
(\ref{Eq:epsparallel}) are (a) $\Delta_1 = 1.495-1.623\cdot10^{14}\,{\rm rad/s}$ and 
$\Delta_2 = 1.778-1.826\cdot10^{14}\,{\rm rad/s}$ and (b) $\Delta_1 = 1.495-1.712\cdot10^{14}\,{\rm rad/s}$ 
and $\Delta_2 = 1.712-1.827\cdot10^{14}\,{\rm rad/s}$.

\begin{figure}[Hhbt]
  \centering
  \epsfig{file=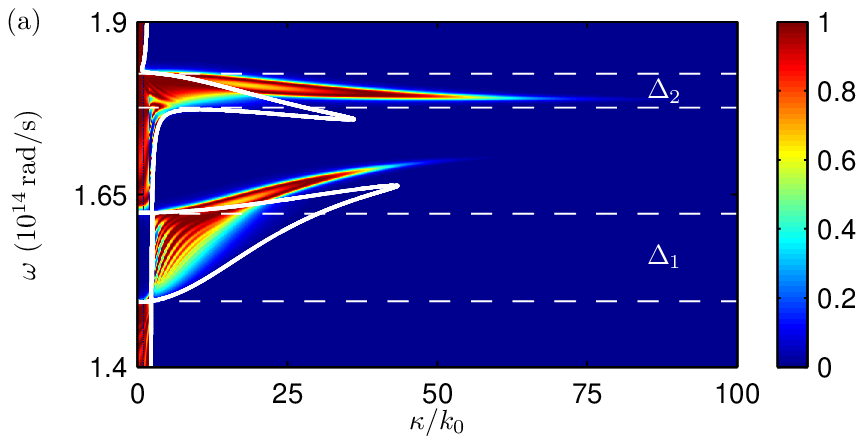}
  \epsfig{file=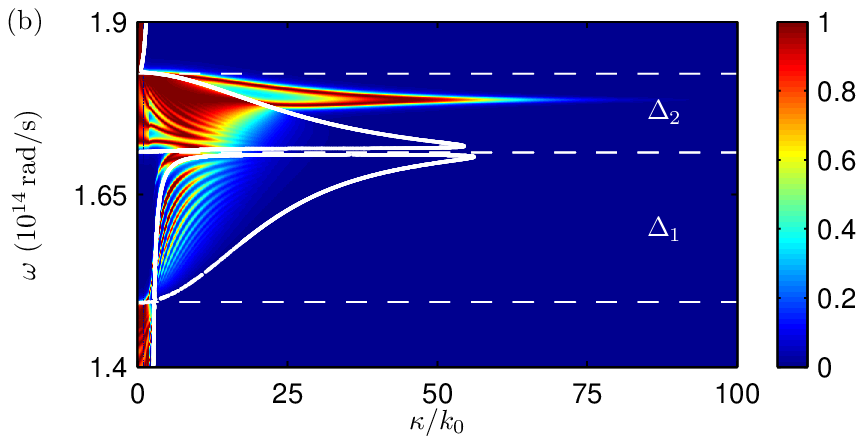}
  \caption{\label{Fig:TransmissionsCoefficient} Transmission coefficient $\mathcal{T}_\rp(\omega,\kappa;d)$ 
           from Eq.~(\ref{Eq:TransmissionCoeff}) for
           both SiC-SiO$_2$ multilayer structures (a) $l_1 = 50\,{\rm nm}$ and $l_2 = 150\,{\rm nm}$, and 
           (b) $l_1 = l_2 = 100\,{\rm nm}$ for the interplate distance $d = 100\,{\rm nm}$.}
\end{figure}

In order to see the structure of contributing modes we have plotted the transmission 
coefficient $\mathcal{T}_\rp(\omega,\kappa;d)$ in Fig.~\ref{Fig:TransmissionsCoefficient}. The horizontal dashed white 
lines mark the hyperbolic bands as determined from effective medium theory~\cite{Yeh}. The solid white 
lines are the borders of the Bloch bands as determined from  Bloch mode dispersion relation for
p polarization~\cite{Yeh}
\begin{equation}
\begin{split}
 \cos(k_{z,B} (l_1 + l_2))&=-\frac{1}{2}\Bigl( \frac{\epsilon_2 k_{z1}}{\epsilon_1 k_{z2}} + \frac{\epsilon_1 k_{z2}}{\epsilon_2 k_{z1}} \Bigr) \sin(k_{z1}l_1) \sin(k_{z2}l_2)  \\
&\quad +\cos(k_{z1}l_1) \cos(k_{z2}l_2),
\end{split}
\label{EQ:BlochDisp}
\end{equation}
with the permittivities $\epsilon_{i}$ ($i = 1,2$) of the two layer materials 
and the wavevector along the optical axis in $z$ direction $k_{zi} = \sqrt{k_0^2 \epsilon_i - \kappa^2}$. Note
that $k_{z,B}$ is the Bloch wavevector inside the multilayer structure which can be approximated by its homogenized
version $k_z$ in Eq.~(\ref{Eq:HyperbolicDisp}) together with Eqs.~(\ref{Eq:epsperp}) and (\ref{Eq:epsparallel}) 
in the regime where the effective description is valid.
Only inside these Bloch bands one can find modes which are propagating modes inside the hyperbolic material. It can 
be seen that there are also very dominant modes outside the Bloch bands contributing significantly to the 
energy transmission. These modes are the coupled surface modes of the topmost SiC layers of each hyperbolic 
material which means that they are evanescent modes inside and outside the hyperbolic structure. 

\begin{figure}[Hhbt]
  \centering
  \epsfig{file=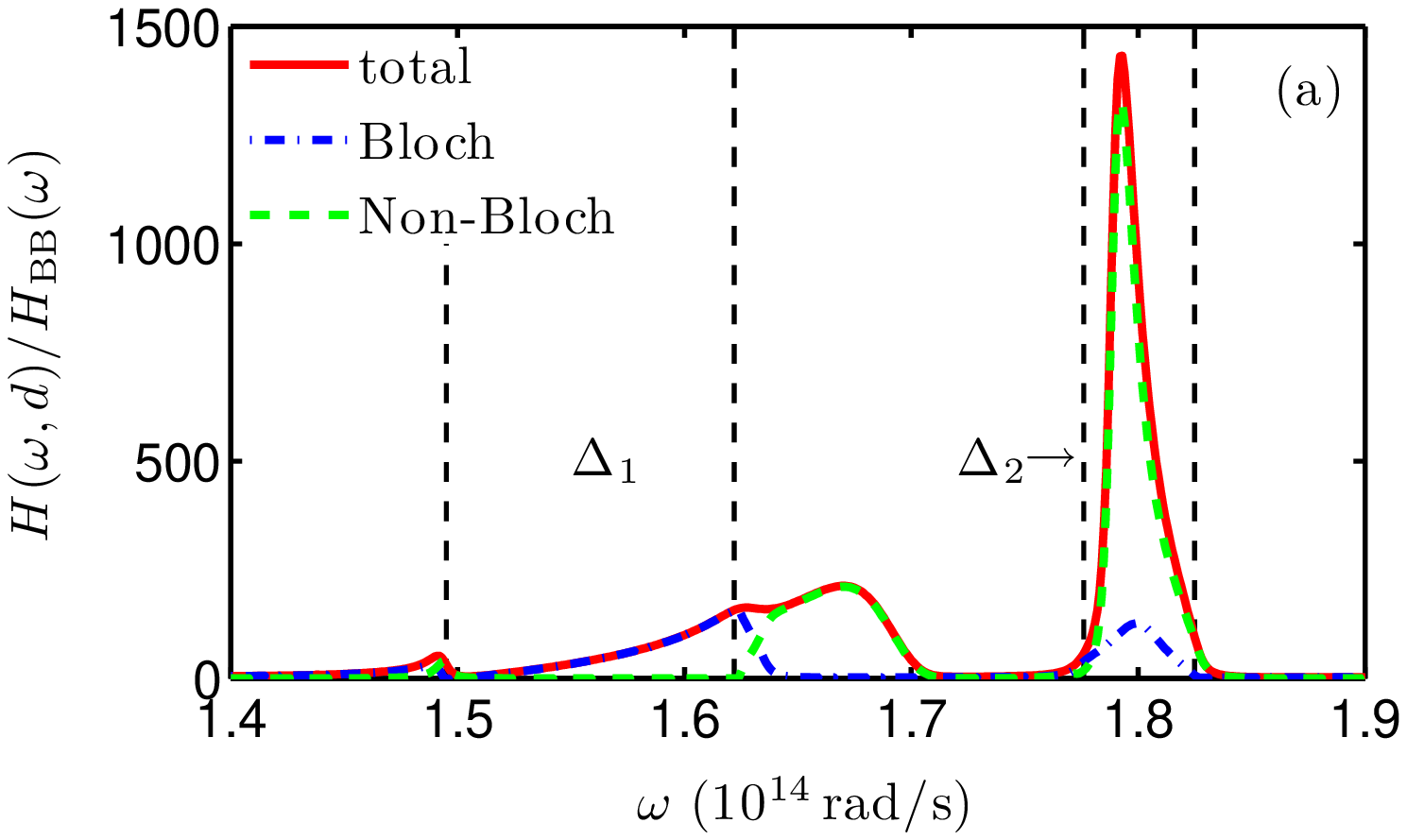, height = 0.3\textwidth}
  \epsfig{file=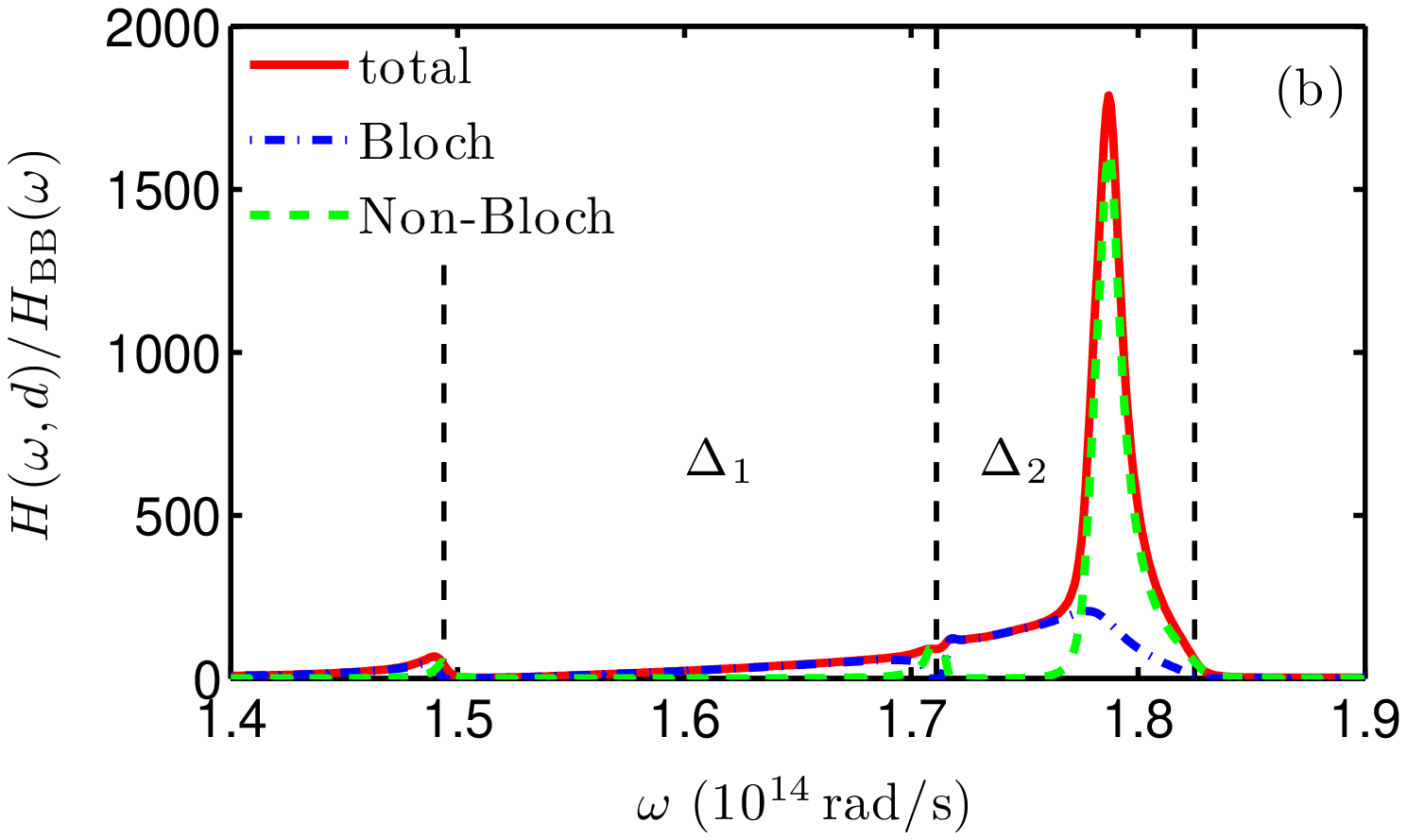, height = 0.3\textwidth}
  \caption{\label{Fig:SpectralHeatFlux} Spectral heat transfer coefficient $H(\omega,d)$ defined in Eq.~(\ref{Eq:htc}) 
           normalized to the black-body result $H_{\rm BB}(\omega) = \omega^2/(2 \pi c^2)$ for
           both SiC-SiO$_2$ multilayer structures (a) $l_1 = 50\,{\rm nm}$ and $l_2 = 150\,{\rm nm}$, and 
           (b) $l_1 = l_2 = 100\,{\rm nm}$ for the interplate distance $d = 100\,{\rm nm}$. Here we choose $T = 300\,{\rm K}$.
           The vertical dashed lines mark the borders of the hyperbolic frequency bands $\Delta_1$ and $\Delta_2$.}
\end{figure}

\begin{figure}[Hhbt]
  \centering
 \epsfig{file=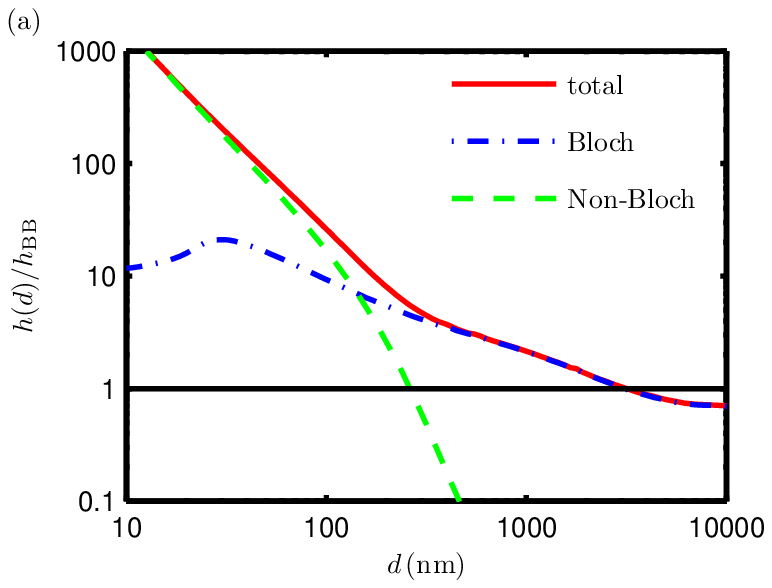}
 \epsfig{file=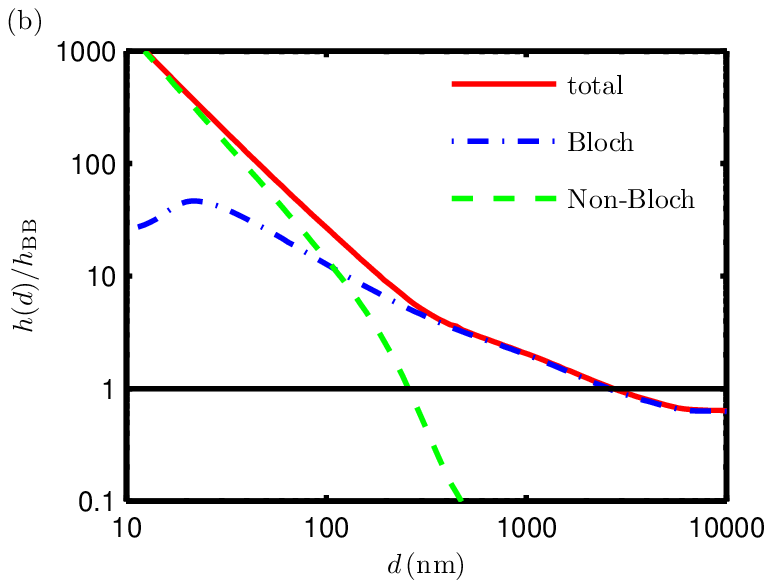}
  \caption{\label{Fig:htc} Heat transfer coefficient $h(d)$ from Eq.~(\ref{Eq:htc}) as a function of interplatedistance $d$
           using $T = 300\,{\rm K}$ for both SiC-SiO$_2$ multilayer structures (a) $l_1 = 50\,{\rm nm}$ and $l_2 = 150\,{\rm nm}$, and (b) $l_1 = l_2 = 100\,{\rm nm}$. The heat transfer coefficient is normalized to the black-body value $h_{\rm BB} = 6.1\,{\rm W}{\rm m}^{-2} {\rm K}^{-1}$. The contributions from the Bloch bands and from regions outside the Bloch bands are shown separately.}
\end{figure}

The respective contribution of the modes inside and outside the Bloch bands to the spectral heat transfer 
coefficient $H(\omega,d)$ is plotted in Fig.~\ref{Fig:SpectralHeatFlux} for a distance of $d = 100\,{\rm nm}$. 
From that figure it becomes apparent that within the hyperbolic frequency bands one has quite large 
contributions steming from modes outside the Bloch bands which are mainly the coupled surface modes of 
the topmost layers. Hence,
for the chosen structure the broadband super-Planckian radiation from the hyperbolic frequency band 
is not due to hyperbolic modes only. The relative contribution of surface modes and all the other modes 
is plotted in Fig.~\ref{Fig:htc} where we show the heat transfer coefficient as a function of distance. From
that figure it becomes obvious that for distances about $100\,{\rm nm}$ and smaller the heat flux is dominated
solely by the coupled surface modes of the topmost layers showing a 
typical $1/d^2$ dependence~\cite{SurfaceScienceReports,Volokitin2007}. 
Whereas for larger distances the heat flux is dominated by the contributions inside the Bloch bands. 
These contributions are on the one hand hyperbolic modes steming from frequencies inside the hyperbolic 
bands $\Delta_1$ and $\Delta_2$. On the other, for frequencies outside the frequency bands $\Delta_1$ 
and $\Delta_2$ the modes are usual propagating or frustrated total internal reflection modes. Note that
for distances of the order of $\max(l_1,l_2) / \pi$ the Bloch-mode contribution reaches a maximum. This
can be attributed to the large wavevector cutoff by the edge of the Bloch bands which can be understood
as the inset of nonlocal effects since for such distances the main wavevector contributions to the thermal 
emission are of the order $\pi/d$.

\begin{figure}[Hhbt]
  \centering
 \epsfig{file=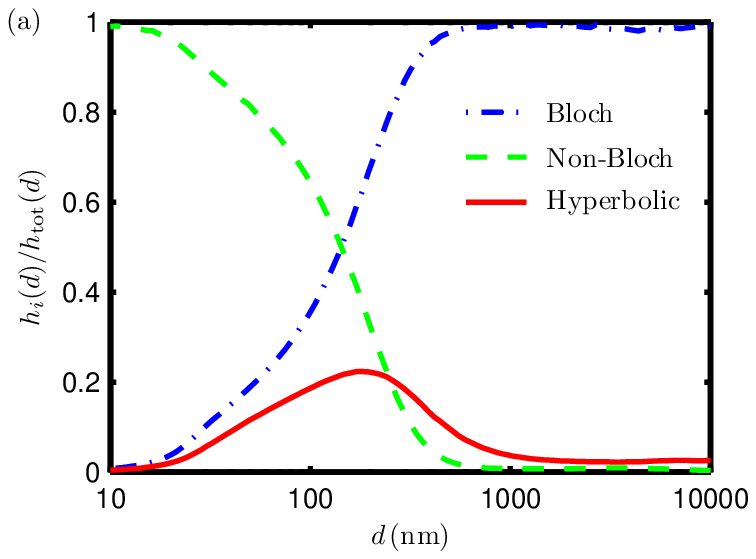}
 \epsfig{file=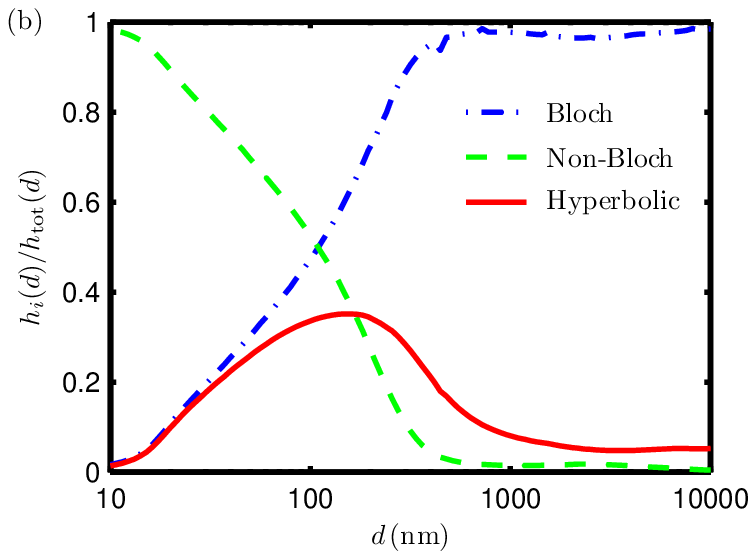}
  \caption{\label{Fig:htc_hm_bloch_surface} Heat transfer coefficients $h_B$, $h_{NB}$, and $h_{hm}$ of the Bloch modes,
           the modes outside the Bloch bands and the hyperbolic modes normalized to the total heat transfer coefficient 
           $h_{tot}(d) = h_B(d) + h_{NB}(d)$ as a function of distance. Again we show both cases (a) $l_1 = 50\,{\rm nm}$ and $l_2 = 150\,{\rm nm}$, and 
           (b) $l_1 = l_2 = 100\,{\rm nm}$, and set $T = 300\,{\rm K}$.}
\end{figure}

To quantify the heat flux mediated by the hyperbolic modes we plot 
in Fig.~\ref{Fig:htc_hm_bloch_surface} the different contributions of the modes inside and outside 
the Bloch bands, and the contribution from the hyperbolic modes separately. The separate contributions 
$h_B(d)$, $h_{NB}(d)$, and $h_{hm}(d)$ to the heat transfer coefficient are normalized to the total heat transfer 
coefficient $h_{tot}(d) = h_B(d) + h_{NB}(d)$. Apparently, in both configurations the contribution of the 
hyperbolic modes is for all chosen distances smaller than $35\%$. This is a rather small value for a 
hyperbolic structure which is constructed for the purpose of enhancing the thermal radiation by the 
hyperbolic-mode contribution.

\begin{figure}[Hhbt]
  \centering
\epsfig{file=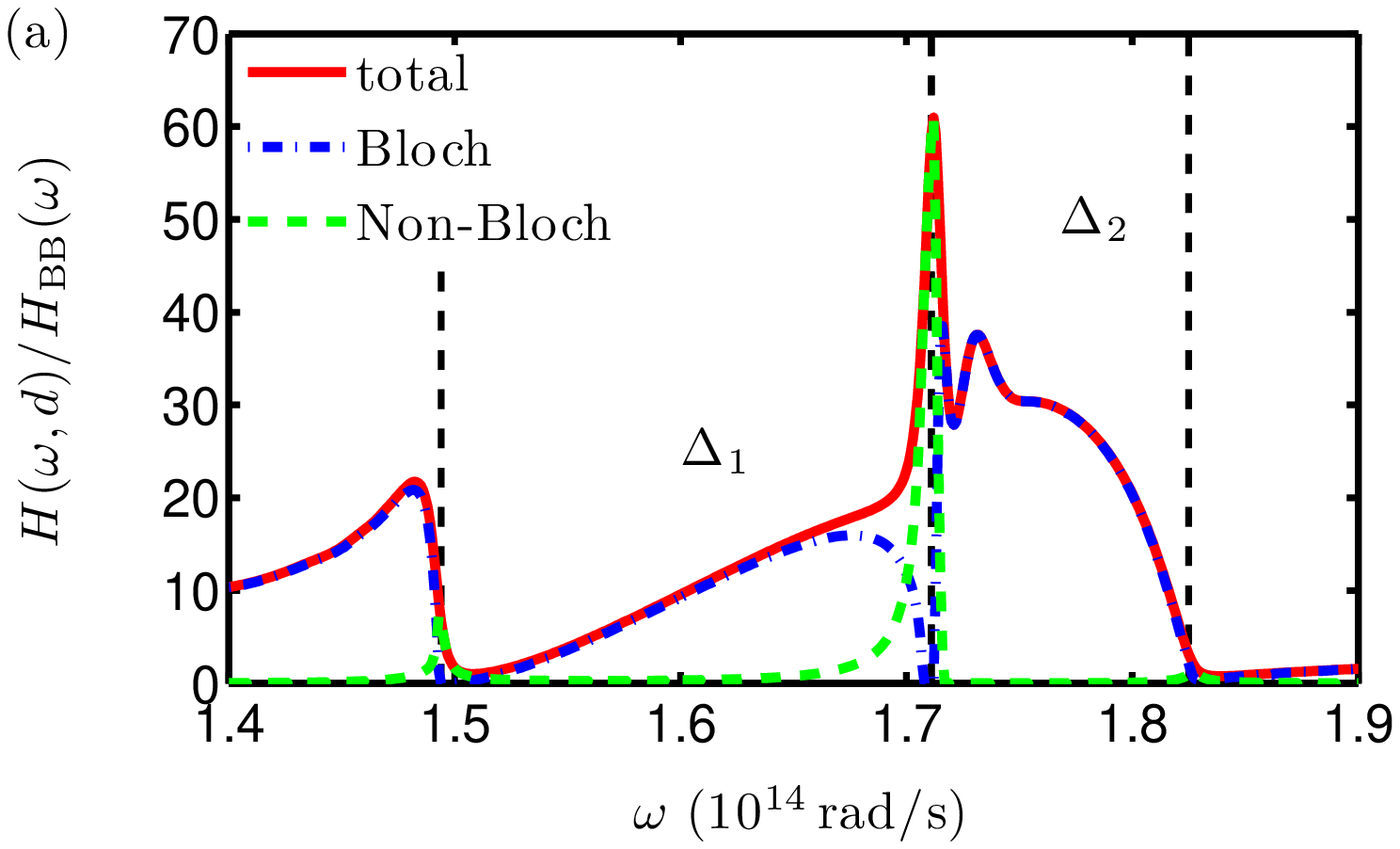, height = 0.3\textwidth}\\
\epsfig{file=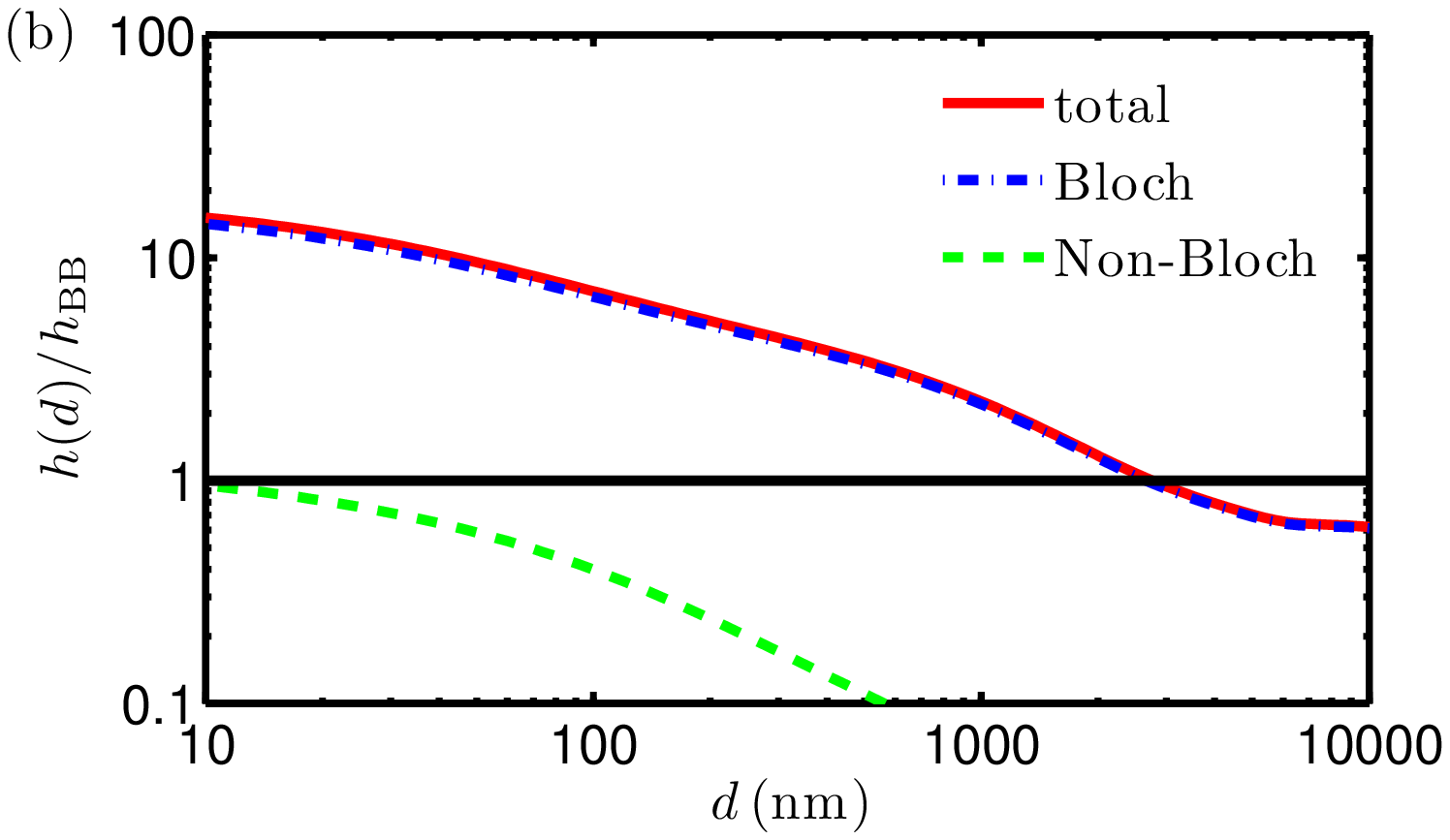, height = 0.3\textwidth}\\
\epsfig{file=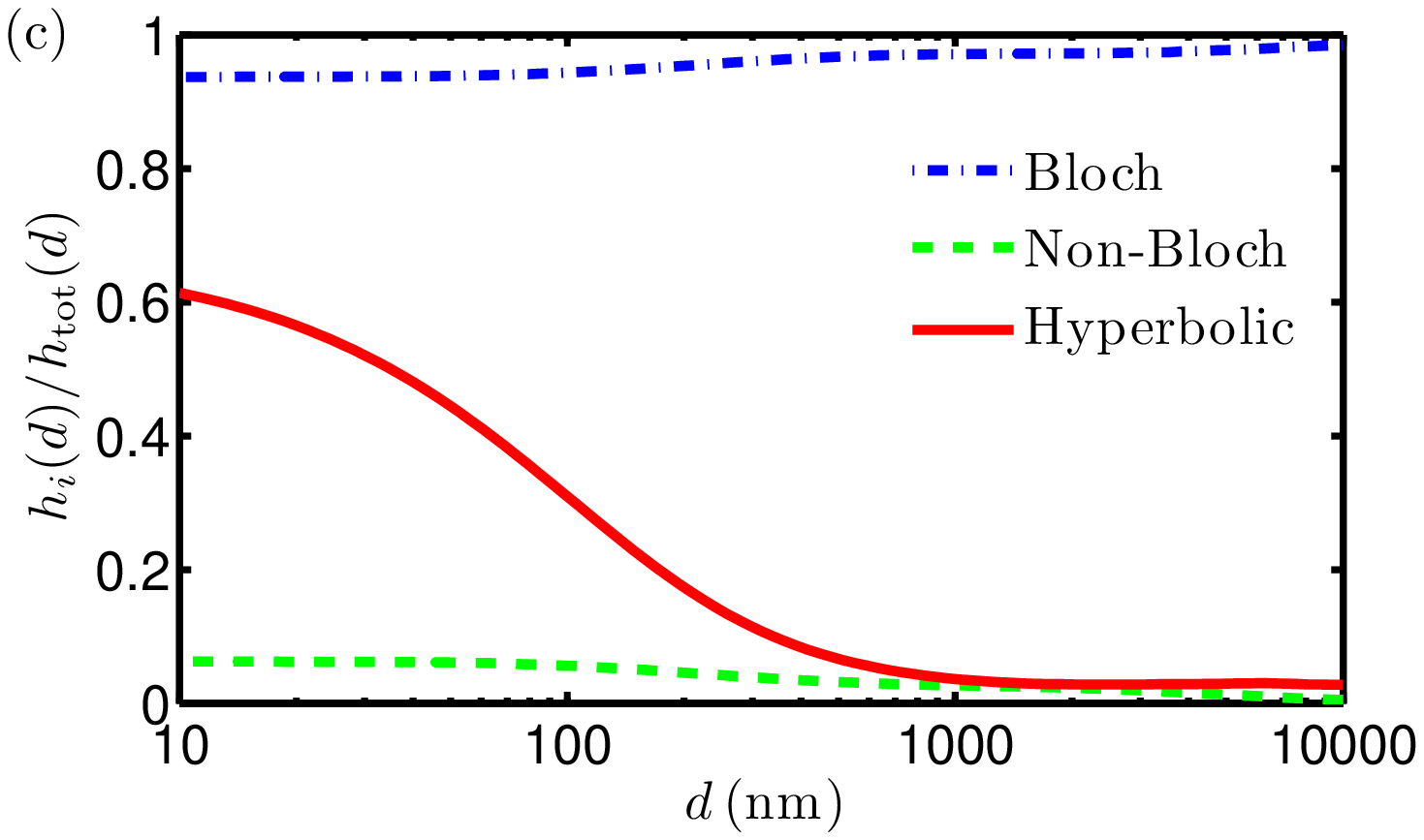, height = 0.3\textwidth}
  \caption{\label{Fig:htc_hm_bloch_surface_inv} (a) Spectral heat transfer coefficients $H(\omega,d)$ between two hyperbolic 
           materials with SiO$_2$ as topmost layer choosing $d = 100\,{\rm nm}$. The vertical dashed lines mark 
           the borders of the hyperbolic frequency bands $\Delta_1$ and $\Delta_2$. (b) Heat transfer coefficients $h(d)$
           for the same materials setting $T = 300\,{\rm K}$ normalized to the black-body value $h_{\rm BB} = 6.1\,{\rm W}{\rm m}^{-2} {\rm K}^{-1}$. Finally in (c) we plot the relative contributions of the Bloch modes, Non-Bloch modes and the hyperbolic modes.}
\end{figure}

Now, let us see if the dominant surface-mode contribution vanishes when choosing the passive SiO$_2$-layer as the topmost
layer. Here, it is important to keep in mind that SiO$_2$ supports surface modes in the infrared. We assume here that it can be
described by a constant permittivity in the frequency band of interest (it is in this sense ''passive'') 
in order to compare our results to existing results in the literature. Hence, we repeat the same calculations 
for the same structure as before but with the difference that for both hyperbolic structures the topmost layer is SiO$_2$ followed by SiC, etc. 
The results for the spectral heat transfer coefficient are shown in Fig.~\ref{Fig:htc_hm_bloch_surface_inv} (a). 
There is still a surface-mode contribution, but it is very small compared to the Bloch-mode contributions. Finally, 
from the distance dependent results in Fig.~\ref{Fig:htc_hm_bloch_surface_inv} (b) and (c) 
it can be seen that the super-Planckian radiation is mainly due to Bloch modes, i.e.\ frustrated total internal 
reflection modes and hyperbolic modes. In particular, the contribution of the hyperbolic modes can be larger 
than 50\% in the strong near-field regime for distances of about $10\,{\rm nm}$. 

\begin{figure}[Hhbt]
  \centering
  \epsfig{file=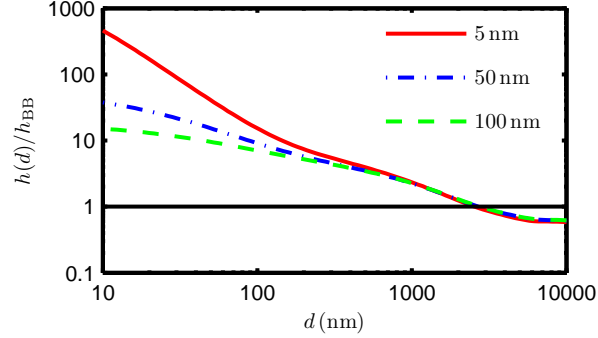}
  \caption{\label{Fig:htc_diff_thickn_inv} The heat transfer coefficient for the structure with the passive material 
           as topmost layer for different layer thicknesses $l_1 = l_2$ ($f=0.5$) of $100\,{\rm nm}$, $50\,{\rm nm}$, and $5\,{\rm nm}$ normalized to the black-body value $h_{\rm BB} = 6.1\,{\rm W}{\rm m}^{-2} {\rm K}^{-1}$.}
\end{figure}

As is obvious from Fig.~\ref{Fig:htc_hm_bloch_surface_inv} (b) the overall heat 
flux is only one order of magnitude larger than that of a black body so that the hyperbolic material considered 
here and in Ref.~\cite{GuoEtAl2012} is a poor near-field emitter compared to the previous structures with SiC 
as topmost layer. But there is a simple method for increasing the hyperbolic contribution by just making
the thickness of the layers smaller. Then, the border of the Bloch bands will shift to larger wavevectors which
results in a broadband contribution to the transmission coefficient for larger wavevectors and hence to 
a larger thermal radiation. In Fig.~\ref{Fig:htc_diff_thickn_inv} we show the heat flux for hyperbolic 
structures with a filling factor of 0.5 but layer thicknesses of $100\,{\rm nm}$, $50\,{\rm nm}$ and $5\,{\rm nm}$. 
It can be seen that the heat flux
increases by orders of magnitude in the strong near-field regimes, i.e., for distances smaller 
than $100\,{\rm nm}$, when making the layers thinner. We have checked that the main contribution is due to 
hyperbolic modes in that regime (not shown here). Further studies have to find an optimized design and 
optimal composite materials in order to further improve the thermal radiation properties of hyperbolic 
materials to attain thermal heat fluxes wich are as large as the heat flux by surface modes or even larger. 
Note that in Ref.~\cite{BiehsEtAl2012} such a structure was proposed on the basis of an effective description. 

In conclusion, we have studied the super-Planckian emission of hyperbolic structures by using 
the fluctuational electrodynamics theory combined with the S-matrix method. It has been shown that 
to properly describe the energy exchanges it is of crucial importance not only to choose a good 
combination of material composites for having broad-band super-Planckian
radiation but also to use a passive material as topmost layer, i.e.\ a material which does not support surface mode
resonances within the thermally accessible spectrum. Also, we have shown for multilayer structures that the thickness
of layers determines the wavevector cutoff of the Bloch band so that it appears clearly advantageous to use thin
layers with elementary thicknesses $l_1,l_2 \ll d$ to observe a large super-Planckian emission at 
a given distance $d$ from the surface. These findings provide the basis for realizing an optimized design 
of hyperbolic thermal emitters with broad-band super-Planckian spectra.

\begin{acknowledgments}
M.T. gratefully acknowledges support from the Stiftung der Metallindustrie im Nord-Westen. The authors acknowledge financial support by the DAAD and Partenariat Hubert Curien Procope Program (project 55923991). This work has been partially supported by the Agence Nationale de la Recherche through the Source-TPV project ANR 2010 BLANC 0928 01. 
\end{acknowledgments}

\end{document}